\documentclass[%
 reprint,
 aps,
]{revtex4-1}

\usepackage{graphicx}
\usepackage{dcolumn}
\usepackage{bm}

\usepackage{natbib}
\begin{document}

\preprint{APS/123-QED}

\title{Deterministic subtraction of single photons based on a coupled single quantum dot-cavity system}
\author{Jinjin Du}
\email{jinjin.du@uwaterloo.ca}
\author{Wenfang Li}
\author{Michal Bajcsy}%
 
\affiliation{%
Institute for Quantum Computing, University of Waterloo,\\
200 University Ave W, Waterloo, ON N2L 3G1, Canada\\
Department of Electrical and Computer Engineering, University of Waterloo,\\ 200 University Ave W, Waterloo,
ON N2L 3G1, Canada }%
\date{\today}

\begin{abstract}
We describe a scheme of deterministic single-photon subtraction in a solid-state system consisting of a charged quantum dot coupled to a bimodal photonic-crystal cavity with a moderate magnetic field applied in a Voigt configuration. We numerically simulate injection of optical pulses into one of the modes of the bimodal cavity and show that the system deterministically transfers one photon into the second cavity mode for input pulses in the form of both Fock states and coherent states.   

\end{abstract}

\pacs{Valid PACS appear here}
\maketitle

Photon subtraction from a light field has been proposed as a tool for generation of various nonclassical states of light with potential for applications in quantum information processing and quantum metrology \citep{Ourjoumtsev2007, Wenger2004, Ourjoumtsev2009, Opatrny2000}. Experimentally, the photon subtraction operation has been used to probe fundamental rules of quantum optics, such as, quantum commutation rules \citep{Parigi2007} or coherent state invariance \citep{Zavatta2008}. Non-deterministically, the single-photon subtraction can be typically implemented with an optical beam splitter of a relatively low reflectivity that taps away a tiny fraction of photons from the incoming light field. In this case, the beam splitter acts as the annihilation operator to remove photons from the initial state of light. This approach has been demonstrated to generate a variety of different non-Gaussian states such as Schr$\ddot o$dinger cat states  \citep{Ourjoumtsev2006} and other novel superposition states \citep{Neergaard-Nielsen2006}. However, the very low success rate in this approach, which also depends on the intensity of the incident light, arises from the nonunitarity of the annihilation operator  \citep{Yoshikawa2017}. Recently,  various different schemes have been proposed to realize deterministic single-photon subtraction \citep{Rosenblum2011, Honer2011, Gea-Banacloche2013, Calsamiglia2001}. In 2015, Rosenblum \textit{et al.} from Dayan's lab at Weizman Institute of Science experimentally demonstrated  such deterministic extraction of a single photon from an optical pulse by coupling a single three-level laser-cooled atom to a micro-resonator \citep{Rosenblum2016}. In that experiment, single photon subtraction from an optical pulse is based on single-photon Raman interaction \citep{Pinotsi2008} realized with near-unity efficiency and the subtracted photon is diverted to another guided mode of the system.

 An on-chip deterministic single-photon subtractor would thus have significance for constructing integrated and miniaturized quantum technology devices. 
Since coupling laser-cooled atoms to on-chip photonic structures, such as demonstrated by Ref. \citep{Thompson2013,Nshii2013}, remains a challenging endeavour, there has always been interest in development of quantum photonics systems based on solid-state emitters. In particular, quantum dots (QDs) coupled to photonic crystal cavities are a well-established platform that benefits from the small mode volume and high-quality factor \citep{Hennessy2007, Yoshle2004, Englund2007} of the cavities, as well as from the straightforward integration with other devices \citep{Faraon2007, Terawaki2012}. For example, generation of non-classical light states in this platform has been demonstrated by filtering the input stream of coherent light using the mechanisms of either photon blockade \citep{Bajcsy2013, Muller2015a, Muller2015, Dory2017} or photon-induced tunneling \citep{Majumdar2012a,Reinhard2012,Faraon2008}. Additionally, electron spins in a charged quantum dot (QD) with multiple energy levels have been successfully used for spin-qubit initialization and manipulation \citep{Press2008, Carter2013}. Overall, this QD-photonic crystal system has shown a great potential for creation of scalable quantum networks and quantum information processing \citep{Faraon2011}.
 
In this paper, we investigate a scheme for deterministic single photon subtraction based on a solid-state cavity quantum electrodynamics (cQED) system, consisting of a charged QD coupled to a bimodal photonic-crystal cavity. We start by numerically calculating the single-photon routing properties of this system, arising from the single-photon Raman interaction process, for a driving light pulse which contains one or two photons. When the charged QD presents a lager quadruplet splitting of the excitonic transitions, single photon routing efficiency approaches unity for feasible parameters of the bimodal photonic-crystal cavity. We further validate the effectiveness of the proposed scheme by considering the system driven by a coherent optical pulse with a relatively large average photon number and present the photon statistics of the outputs from the two modes of the bimodal cavity. Our calculations indicate that a deterministic single photon subtraction from an input optical pulse can be implemented with unity efficiency in the proposed system.

\begin{figure}[htb]
\centerline{
\includegraphics[width=9.5cm]{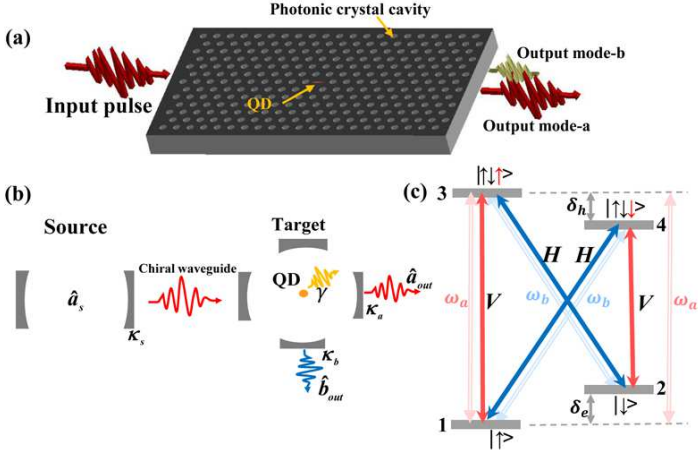}}
 \caption{(Color online) (a) Schematic of the proposed solid-state system for implementing single photons subtraction based on a charged quantum-dot (QD) coupled to a photonic crystals cavity existing two different cavity modes (mode-a and b). (b) Schematic depiction of the studied system is modeled by a cascaded quantum system. A single-sided source cavity as a photon source is used to generate input pulses to resonantly couple to the target system via a chiral waveguide. The emission of the source cavity is fed into the one quantum dot (QD) contained in the target bimodal photonic crystal cavity.
  (c) Energy level configuration for the single charged QD at magnetic fields.}
\end{figure}

Our proposed solid-state system is based on a charged QD embedded in a bimodal photonic-crystal cavity and is schematically illustrated in Fig.1(a). For computational purposes, we consider the system as a cascaded quantum system, shown in Fig.1(b), consisting of a source cavity for producing a well-defined incoming pulse and a target cavity in which the deterministic subtracting single photon from the incoming pulse takes place. Here, the source cavity is defined as a single-sided optical cavity with decay rate ${\kappa _s}$ and the target cavity is a bimodal (i.e., supporting two modes of orthogonal polarization) photonic-crystal cavity such as \textit{H}1 cavity described in Ref. \citep{Fu:13, Coles2014} containing a charged QD coupled to its modes. Note though that other cavity designs, such as that described in Ref. \citep{Rivoire2011}, could potentially also be used. The initial state of the source cavity determines the photon state of the incoming optical pulse, while the pulse width is determined by ${\kappa _s}$. This optical pulse is then unidirectional fed via a link formed by a chiral waveguide into the target system where it couples to one mode of the bimodal cavity. The resulting optical fields arising from the interaction of the charged QD and the two modes of the target cavity are then assumed to escape from one side of the target cavity. Practically, this could be achieved through an asymmetric design of the photonic crystal such as reported for the nanobeam cavities in Ref. \citep{Tiecke2014}.   
Experimentally, Majumdar \textit{et al}. have successfully fabricated this kind of photonic-crystal bimodal cavities \citep{Majumdar20122} and a deterministic charged QD embedded in photonic crystal nanoresonator has been demonstrated in Ref. \citep{Lagoudakis2013}. 


Charged quantum dots, such as InAs dots embedded in GaAs, present four optically active transitions with different energies when an external magnetic field is applied in the Voigt configuration, as illustrated in Fig. 1(c). For the positively-charged dot, energy splitting ${\delta _{h(e)}}$  of ground (excited) states is solely determined by the Land\'{e} g-factor of the electron (hole) as $\delta _{h(e)} = \mu _b g_{h(e)}B$, where ${\mu _b}$ and $B$ denote the Bohr magneton and the magnetic field, respectively. In an experiment, energy  splitting $\delta _{h(e)}\approx $ 30 GHz was observed with magnetic field $B$ = 5T \citep{Lagoudakis2014}. As a result, the charged quantum dot is transformed into a double- $\Lambda$ system with the characteristic signature of a quadruple spectral line. The two outer and inner transitions are the result of the QD coupling to optical fields of different polarizations, either horizontal or vertical, which is determined by transition selection rules \citep{Bayer2002}. Therefore, when one mode (namely mode-\textit{a}) of the bimodal photonic-crystal cavity is excited by a light pulse in vertical (\textit{V}) polarization, both transitions $\left|  \uparrow  \right\rangle  \to \left| { \uparrow  \downarrow  \uparrow } \right\rangle $  and  $\left|  \downarrow  \right\rangle  \to \left| { \uparrow  \downarrow  \downarrow } \right\rangle $ could be coupled with  rate of ${g_a}$. Similarly, the other mode \textit{b} of horizontal (\textit{H}) polarization is permitted to couple the transitions $\left|  \uparrow  \right\rangle  \to \left| { \uparrow  \downarrow  \downarrow } \right\rangle$ and $\left|  \downarrow  \right\rangle  \to \left| { \uparrow  \downarrow  \uparrow } \right\rangle$ with its coupling rate of ${g_b}$. We consider the case that mode-\textit{b} is resonant with two inner transitions by adjusting the detuning of ${\delta _h} \approx {\delta _e} = \delta$. According to the input-output formalism \citep{Gardiner1985}, the output modes of the bimodal photonic-crystal cavity for its two modes (mode-\textit{a} and \textit{b}) are given by

\begin{equation}
\begin{array}{l}
{{\hat a}_{out}} = \sqrt {{{\kappa _s}}} {{\hat a}_s} + \sqrt {{\kappa _a}} \hat a\ ,\\
{{\hat b}_{out}} = \sqrt {{\kappa _b}} \hat b\ ,
\end{array}
\label{eq:refname1}
\end{equation}
where ${\hat a}$, ${\hat b}$ and ${{\hat a}_s}$   are annihilation operators of mode-\textit{a}, mode-\textit{b} of the bimodal photonic-crystal cavity and the source cavity, respectively; ${\kappa _a}$ is the field decay rate of mode-\textit{a} and ${\kappa _b}$ is for that of mode-\textit{b}. Here, we apply quantum trajectory approach \citep{1993LNPM...18.....C} to calculate time evolution of the system, which is described by the Schr$\ddot o$dinger equation  

\begin{equation}
\frac{d}{{dt}}\left| {\psi (t)} \right\rangle  = \frac{1}{{i\hbar }}{H_{eff}}\left| {\psi (t)} \right\rangle\ ,
\end{equation}
with the non-Hermitian Hamiltonian

\begin{equation}
{H_{eff}} = {H_s} + {H_t} + {H_{st}} - \frac{{i\hbar }}{2}{\sum\limits_k {{{\hat C}_k}} ^\dag }{\hat C_k}\ ,
\end{equation}
where ${H_s}$  (${H_t}$) represents the source (target) Hamiltonian; ${H_{st}}$ is the interaction Hamiltonian, and ${\hat C_k}$ is the collapse operator of the system given by

\begin{equation}
\begin{array}{l}
{{\hat C}_1} = {{\hat a}_{out}}\ ,\\
{{\hat C}_2} = {{\hat b}_{out}}\ ,\\
{{\hat C}_3} = \sqrt {{\gamma _{}}} {{\hat \sigma }_{33}}\ ,\\
{{\hat C}_4} = \sqrt {{\gamma _{}}} {{\hat \sigma }_{44}}\ ,
\end{array}
\end{equation}
where ${\sigma _{ij}} = \left| i \right\rangle \left\langle j \right|$ denotes the projection ($i = j$ ) and lowering or rising operator ($i \ne j$ ), respectively. $\gamma $ is defined as spontaneous emission decay rate of the QD. By substituting collapse operators in Eq.(4) to the Eq.(3), the non-Hermitian Hamiltonian ${H_{eff}}$  can be expressed as ($\hbar  = 1$)

 \begin{equation}
 \begin{array}{c}
{H_{eff}} = {g_a}(\hat a{\hat \sigma _{31}} + {{\hat a}^\dag}{\hat\sigma _{13}}) + {g_b}(\hat b{\hat\sigma _{32}} + {{\hat b}^\dag }{\hat\sigma _{23}})\\
{\rm{ }} + {g_a}(\hat a{\hat\sigma _{42}} + {{\hat a}^ \dag }{\hat\sigma _{24}}) + {g_b}(\hat b{\hat\sigma _{41}} + {{\hat b}^ \dag}{\hat\sigma _{14}})\\
{\rm{ }} - i(\frac{\gamma }{2} + i{\omega _{13}}){\hat\sigma _{33}} - i(\frac{\gamma }{2} + i{\omega _{14}}){\hat\sigma _{44}} + {\omega _{12}}{\hat\sigma _{22}}\\
{\rm{ }} - i(\frac{{{\kappa _a}}}{2} + i\omega {}_a){{\hat a}^\dag }\hat a - i(\frac{{{\kappa _b}}}{2} + i{\omega _b}){{\hat b}^ \dag }\hat b\\ - i(\frac{{{\kappa _s}}}{2} + i{\omega _s}){\hat a_s^\dag }\hat a_s- i\sqrt {{\kappa _a}{\kappa _s}} {{\hat a}^ \dag }\hat a_s\ .
\end{array}
\end{equation}
where $\omega _s$, $\omega _a$ and $\omega _b$ are the resonance frequencies of source cavity, mode-a and mode-b of the bimodal photonic-crystal cavity, respectively.

\begin{figure}[htb]
\centerline{
\includegraphics[width=7.5cm]{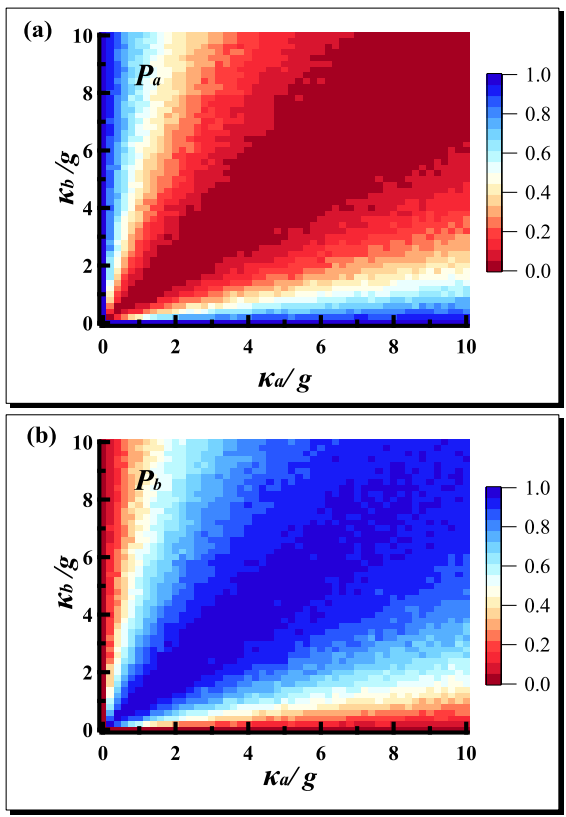}}
 \caption{(Color online) (a)  Intensity plot of photon detection probability   at one mode(mode-a) of the output modes of the bimodal cavity as functions of the cavity decays $\frac{{\kappa _a}}{g}$   and $\frac{{\kappa _b}}{g}$ .(b) the same for photon detection probability   at the other (mode-b) mode; Here we set $g/2\pi= g_a/2\pi  = g_b/2\pi= 10$ GHz,  $\kappa_s/2\pi=50$ MHz$, $ and $\gamma/{2\pi}=0.25$ GHz.}
\end{figure}

To explore the dynamics of the quantum system described by Eq. (2), we use the Quantum Toolbox in Python (QuTiP) \citep{qutip}. First, we consider driving the target cavity by an optical pulse produced by the source cavity with an initial intracavity photon number of one, assuming a typical coupling strength between the QD and the photonic-crystal cavity of $g/2\pi \approx 1 - 30$ GHz \citep{Majumdar20122,Rundquist2014} and mode volumes on the order of ${(\lambda /n)^3}$. 
We investigate the interaction regimes ranging from weak to strong in our solid-state cQED system by sweeping the ratio of $\kappa /g$ \citep{Faraon2008}. Initially, we assume the coupling strength for both modes of the photonic crystal cavity to be identical, with values of $g/2\pi= g_a/2\pi  = g_b/2\pi= 10$ GHz. The initial state of the charged QD is $\left|  \uparrow  \right\rangle $ and the whole system is on-resonance ( $\omega _s= \omega _a = \omega _{13}$, $\omega _b = \omega _{14}$, and $\omega _{12}=\delta$). To achieve efficient single-photon Raman process, a long input driving pulse given by $\kappa_s $ (e.g., a typical value $\kappa_s/2\pi$= 50  MHz) can be produced by the source cavity \citep{Pinotsi2008}.   Figs. 2(a) and (b) show that the dependence of single-photon detection probability for the output modes (mode-\textit{a} and \textit{b}) on the decay rates $\kappa _a$ and $\kappa _b$, respectively. An efficient single-photon Raman interaction can be achieved (${P_a} \to 0,{P_b} \to 1$) in this scheme for $\kappa _a \approx \kappa _b = \kappa$ and $g^2/(\kappa \gamma ) \gg 1$. In other words, the process of single-photon 'transfer' within the system can be summarized as follows: the QD in the bimodal cavity absorbs an incident \textit{V}-polarized photon and then releases a \textit{H}-polarized photon after QD-cavity interaction and the QD state is transferred from $\left|  \uparrow  \right\rangle $ to $\left|  \downarrow  \right\rangle $. In an effort to account for experimental imperfections, such as cavity fabrication, QD location inside the cavity, and QD properties, we further calculate the dependence of photon detection probabilities at the mode-\textit{b} of the cavity output on additional parameters, such as the coupling strength g (illustrated in Fig. 3(a)) and the decay rate of the QD excited state $\gamma $ (shown in Fig.3(b)). The probabilities for all varying parameters can still reach $\sim$0.8 even for $\kappa/2\pi $= 100 GHz. Thus, it can be seen that single-photon Raman interaction in the system is robust 
for a large range experimental parameters.

\begin{figure}[htb]
\centerline{
\includegraphics[width=7.5cm]{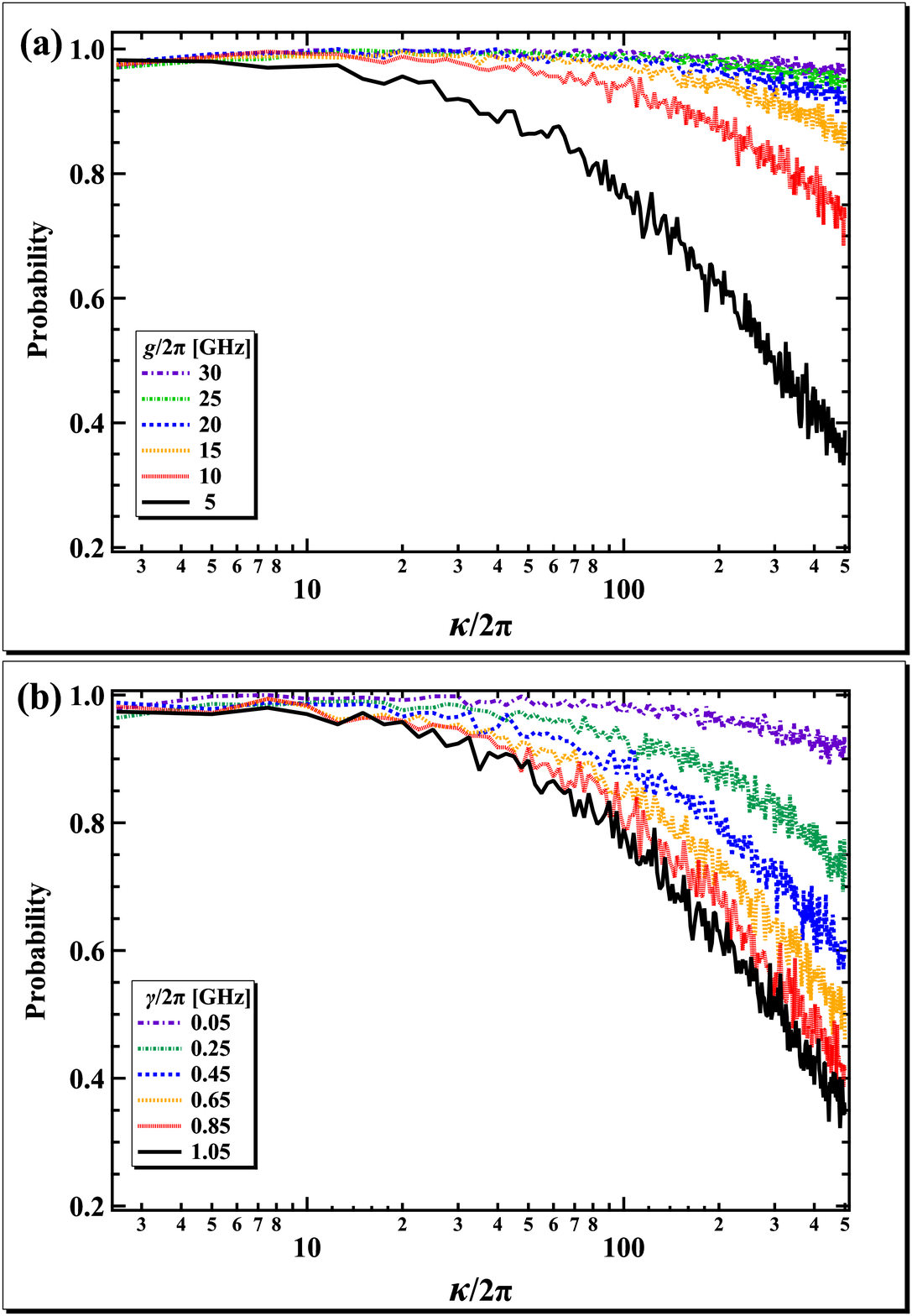}}
 \caption{(Color online) Photon detection probabilities at the output mode-b dependence of $\kappa/ 2\pi$ for varying various (a) $g/2\pi$  and (b) $\gamma/2\pi$ on a log scale. Different curves in (a) correspond to different values $g/2\pi$ from 5 to 30 GHz, with a step of 5 GHz .Curves is shown in (b) for different values $\gamma/2\pi$ increasing from 0.05 to 1.05 GHz with a 0.2-GHz interval. Other parameters are as follows: $g/2\pi= g_a/2\pi = g_b/2\pi$, $\kappa_s/2\pi=50$ MHz,  and $\kappa/{2\pi}=\kappa_a/{2\pi}=\kappa_b/{2\pi}$.}

\end{figure}

\begin{figure}[htb]
\centerline{
\includegraphics[width=10cm]{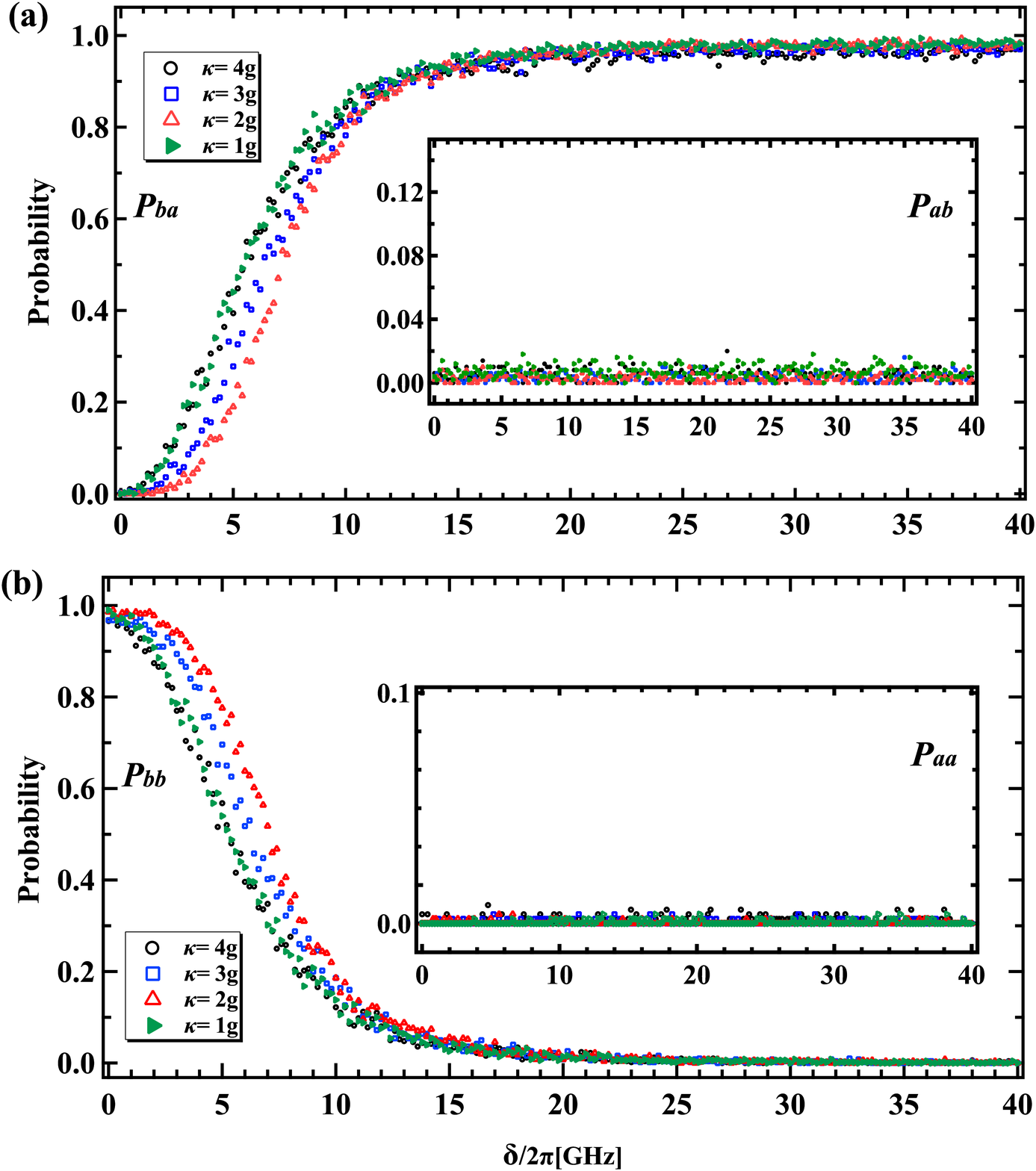}}
 \caption{ (Color online) Two-photon detection probabilities (${P_{ba}}$, ${P_{ab}}$, ${P_{bb}}$ and ${P_{aa}}$) as a function of the Zeeman splitting frequency $\delta/2\pi $ for different $\kappa $ = 1$g $ (Green right triangle), 2$g $ (Red triangle), 3$g $ (Blue block) and  4$g $ (Black circle). (a) denotes  ${P_{ba}}$ and (b) for ${P_{bb}}$. The corresponding insets show ${P_{ab}}$  and ${P_{aa}}$, respectively. Other parameters are set as in Fig.2.}
 
\end{figure}

To further investigate the performance and characteristics of the operation of single-photon subtraction, we consider photon routing properties of the system under the same condition as above, except with the input driving pulse containing two photons in order to show that efficiency of photon routing approaching unity is also achievable for the input pulse containing multiphoton Fock states. Conceptually, in this case, the first photon transfers the QD from the level $\left|  \uparrow  \right\rangle  \to \left|  \downarrow  \right\rangle $ and, at the same time, a \textit{H}-polarized photon is released in the mode-\textit{b}, which results in  the QD no longer interacting with the incoming \textit{V}-polarized photon in mode-\textit{a} if there is sufficiently large detuning between mode-\textit{a} and the transition  $\left|  \downarrow  \right\rangle  \to \left| { \uparrow  \downarrow  \downarrow } \right\rangle $.
In more detail, this scenario results in four possible detection events for output pulses: $P_{ba}$ is for detection a \textit{V}-polarized photon from the mode-\textit{a} following detection of a \textit{H}-polarized photon from the mode-\textit{b}; $P_{bb}$ for both photons from mode-\textit{b} with \textit{H}-polarization; $P_{ab}$ represents detecting a \textit{H}-polarized photon from the mode-b after detecting a \textit{V}-polarized photon from the mode-\textit{a}; $P_{aa}$ denotes both photons escaping from mode-\textit{a} with \textit{V}-polarization. The photon routing efficiency $P_{c}$ is then defined as $P_c = P_{ab} + P_{ba}$.

Figures 4(a) and (b) and their insets show the two-photon detection probability ($P_{ba}$,  $P_{bb}$ , $P_{ab}$ and $P_{aa}$) as a function of the Zeeman splitting $\delta $ of the ground and excited states, respectively. Note that it is experimentally possible to choose quantum dots with almost the same Zeeman splitting $\delta $ for the ground state and excited states \citep{Lagoudakis2014}. When the $\delta $ is equal to zero as shown in Fig.4 (b), all the photons escape from the mode-b of the bimodal cavity (${P_{bb}} \approx 1$) because the system can be considered as a double $\Lambda$ scheme to implement two single-photon Raman interaction processes. After the two steps of single-photon Raman interaction, each photon of the input pulse transfers to mode-\textit{b},where it is detected, and the QD state is reverted to its initial state $\left|  \uparrow  \right\rangle $. Under this condition, the system can be used to change the polarization of  incident beam of light from \textit{V(H)} to \textit{H(V)} polarization rather than as an efficient photon router. From Figs. 4(a) and (b), we note that ${P_{ba}} \to 1,{P_{bb}} \to 0$ with increase in the Zeeman splitting frequency $\delta $ , which leads to the second photon no longer interacting with the transition $\left|  \downarrow  \right\rangle  \to \left| { \uparrow  \downarrow  \downarrow } \right\rangle $. The inset figures  show that the probability of events for ${P_{ab}}$ and ${P_{aa}}$ approach zero. Therefore, photon routing efficiency of ${P_c} \approx {P_{ba}}$ approaches unity with splitting frequency $\delta /2\pi  > 20$ GHz -- a value which can be achieved in existing experiments with a few Tesla magnetic field \citep{Lagoudakis2013, Lagoudakis2014}. We also note that the two-photon detection probabilities for different $\kappa $ have good robustness (see Fig. 4).

Next, we simulate single-photon subtraction when the system is driven by a resonant pulsed coherent light (${\omega _s} = {\omega _a} = {\omega _{13}}$, ${\omega _b} = {\omega _{14}}$, and $\omega _{12}=\delta$). We choose the system parameters as $g/2\pi=10$ GHz, $\gamma/{2\pi}=0.25$ GHz, $\kappa/2\pi=20$ GHz, and $\delta/2\pi=25$ GHz, which correspond to parameters commonly reported for this platform in the literature. 
Fig. 5 shows the detected mean photon number ${{\bar n}_{Out}}$ with either \textit{V}- polarization from the mode-a or \textit{H}-polarized photon from mode-\textit{b} as a function of the mean input photon number ${{\bar n}_{In}}$. The mean photon number in \textit{H}-polarization from the mode-\textit{b} remains a constant equal to 1, as displayed by  the blue circle line in Fig. 5. 
Similarly, the \textit{V}-polarization photon from mode-\textit{a} is displayed by the red solid-circle line for its coupling to the QD and the black circle line for its non-coupling to the QD. It reads that one photon is determinately subtracted from the input coherent state pulse and transferred to the mode-\textit{b} of the bimodal photonic crystal cavity.

\begin{figure}[htb]
\centerline{
\includegraphics[width=7.5cm]{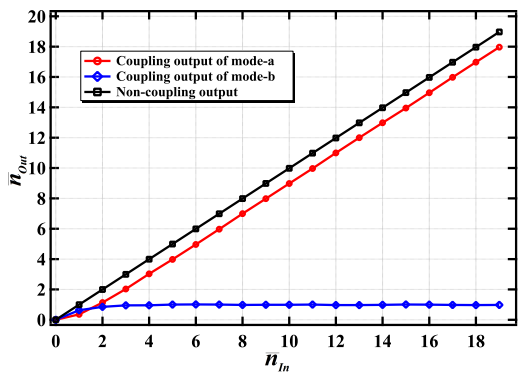}}
 \caption{(Color online) Average photon number ${{\bar n}_{Out}}$ for the output mode-a (red circle line) and mode-b photons (blue circle line) verse the average photon number ${\bar n_{In}}$ of the input pulse. The red circle curve is displayed for the photon interacting with QD and the black circle line is for the case of the photon non-coupling with QD.}
\end{figure}

\begin{figure}[htb]
\centerline{
\includegraphics[width=7.5cm]{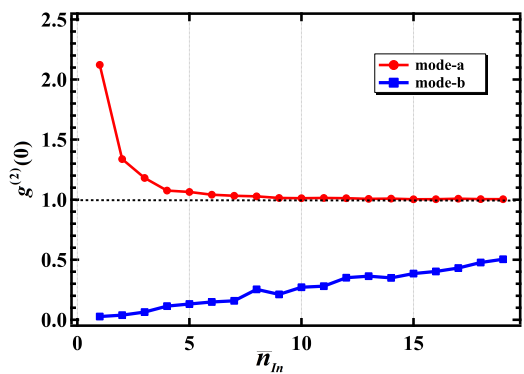}}
 \caption{(Color online) Numerical simulation of the second order correlation function $g^{(2)}(0)$ as a function of the average photon number ${\bar n_{In}}$ of the input pulse. Red circle line: $g^{(2)}(0)$ for the output mode of mode-a ; Blue block line: $g^{(2)}(0)$ for the output field of mode-b.}
\end{figure}

\begin{figure}[htb]
\centerline{
\includegraphics[width=8cm]{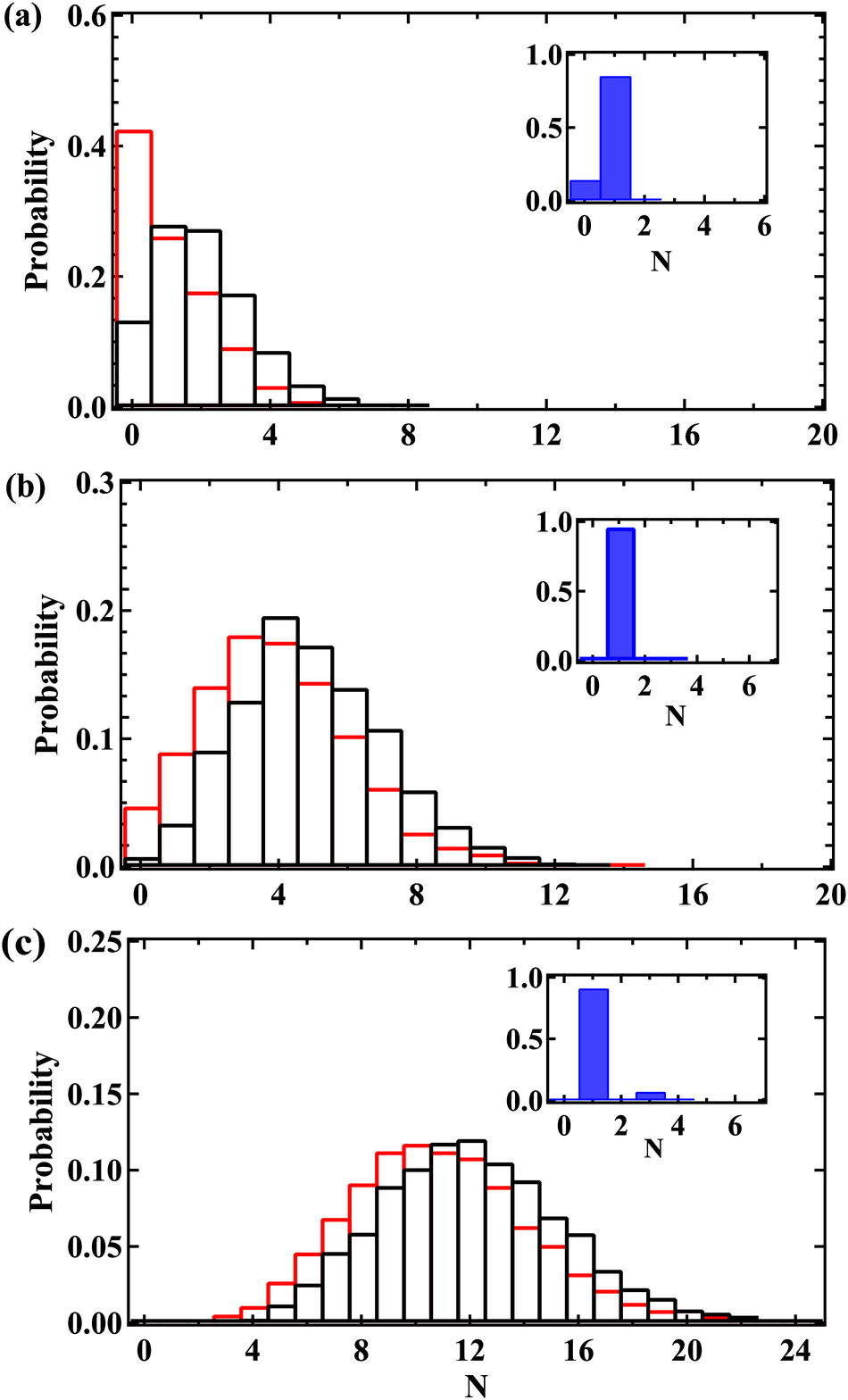}}
 \caption{(Color online) Photon statistics of the output field of  the mode-a with the QD in the target system compared with the initial field without QD inside. The average input photon number ${{\bar n}_{in}}$ are 2, 5 and 12 for (a), (b) and (c) respectively. Red bar: statistics distribution of the subtracted single-photon field from mode-\textit{a}; Black bar: statistics distribution of the output field of the mode-a without QD. The inset figures (Blue bar) are the corresponding the statistics of the output mode of mode-\textit{b}.}
\end{figure}

To evaluate the photon statistics of the subtracted photon state escaping from the cavity mode-\textit{b}, we calculate the second-order correlation function $g^{(2)}(0)$ as a function of ${{\bar n}_{In}}$, as shown in Fig. 6. Here, $g^{(2)}(0)$ of the light in output of mode-\textit{b} remains well below the classical limit for the simulated range of ${{\bar n}_{In}}$, which shows that the system indeed extracts a single-photon Fock state from the coherent input field. 
At the same time, the output of mode-\textit{a} displays photon bunching ($g^{(2)}(0)>1$) in its statistics for low values of ${\bar n}_{In}$ with the values of $g^{(2)}(0)$ decreasing back to the classical limit as the ${\bar n}_{In}$ increases. This phenomenon can be explained by considering that removing a single photon from a weak coherent pulse results in a photon state that can be described as $\psi \approx \sqrt{1-\varepsilon^2}\vert 0\rangle + \varepsilon \vert 2\rangle + ...$, with $\varepsilon \ll 1$, for which $g^{(2)}(0)\approx \frac{1}{2\varepsilon^2}$ \citep{Rundquist2014}. As ${\bar n}_{In}$ increases, this approximation no longer holds and the effect of the single photon removal on $g^{(2)}(0)$ of the output of mode-\textit{a} fades away.
 The deterministic single-photon subtraction can be further confirmed by plotting the actual photon-number distribution in the output optical fields of the mode-\textit{a} and mode-\textit{b}, shown in Fig.7.  The initial coherent-state driving pulses have a typical Poissonian distribution of photon numbers with the average photon number ${\bar n}$, shown for ${\bar n}$ of 2, 5 and 12 in Fig.7 (black bars). The red bars in the figure then show the photon number distribution in the output of mode-\textit{a} after the input pulse interacted with our photon subtracting QD-cavity system. The envelope of the photon number distribution is shifted downwards compared to the input pulse, with a new average photon number of $(\bar n - 1)$. The calculated photon distribution of the output of mode-\textit{b} is presented by the insets in Fig.7, showing this output to be a dominantly single-photon pulse.

In conclusion, we presented a scheme for deterministic single-photon subtraction based on a solid-state cQED system consisting of a single charged QD coupled to a bimodal photonic-crystal cavity. We used quantum trajectory approach to numerically simulate this system and explored the performance of the proposed scheme using a source cavity to input selected photon states into the system. Our simulation results suggest that deterministic single photon subtraction from various input optical pulses can be realized with parameters available in current experimental systems. The capability of our proposed on-chip solid-state device to deterministically subtract single photons from an arbitrary input pulse makes it a promising tool for scalable applications in quantum state engineering and quantum information processing, ranging from generation of nonclassical states of light \citep{Ourjoumtsev2006} to implementing a photon-number-resolving detector by integrating and concatenating several of these devices \citep{2016NaPho..10....4B}.

\begin{acknowledgments}
This research was undertaken thanks in part to funding from the Canada First Research Excellence Fund. We would like to thank Sreesh Venuturumilli, Behrooz Semnani, and Arka Majumdar for helpful discussions.
\end{acknowledgments}

\nocite{*}

\bibliography{references}

\end{document}